\documentclass[prd,preprint,nofootinbib,superscriptaddress]{revtex4-1}
\usepackage{amsmath}
\usepackage{latexsym}
\usepackage{amsfonts}
\usepackage{graphicx}
\usepackage{mathrsfs}
\usepackage{CJK}
\usepackage{subfigure}
\usepackage{float}
\usepackage{color}
\usepackage{hyperref}
\hypersetup{
    colorlinks=true,
    linkcolor=red,
    citecolor=blue,
}

\begin{document}
\begin{CJK*}{GB}{gbsn}

\title{Frequency response of space-based interferometric gravitational-wave detectors}
\author{Dicong Liang}
\email{dcliang@hust.edu.cn}
\affiliation{School of Physics, Huazhong University of Science and Technology,
Wuhan, Hubei 430074, China}

\author{Yungui Gong}
\email{Corresponding author. yggong@hust.edu.cn}
\affiliation{School of Physics, Huazhong University of Science and Technology,
Wuhan, Hubei 430074, China}

\author{Alan J. Weinstein}
\email{ajw@ligo.caltech.edu}
\affiliation{LIGO Laboratory, California Institute of Technology, Pasadena, California 91125, USA }

\author{Chao Zhang}
\email{chao\_zhang@hust.edu.cn}
\affiliation{School of Physics, Huazhong University of Science and Technology,
Wuhan, Hubei 430074, China}

\author{Chunyu Zhang}
\email{chunyuzhang@hust.edu.cn}
\affiliation{School of Physics, Huazhong University of Science and Technology,
Wuhan, Hubei 430074, China}

\begin{abstract}
Gravitational waves are perturbations of the metric of space-time. Six polarizations are possible, although general relativity predicts that only two such polarizations, tensor plus and tensor cross are present for gravitational waves. We give the analytical formulas for the antenna response functions for the six polarizations which are valid for any equal-arm interferometric gravitational-wave detectors without optical cavities in the arms.
The response function averaged over the source direction and polarization angle decreases at high frequencies
which deteriorates the signal-to-noise ratio registered in the detector.
At high frequencies, the averaged response functions for the tensor and breathing modes fall of as $1/f^2$,
the averaged response function for the longitudinal mode falls off as $1/f$ and
the averaged response function for the vector mode falls off as $\ln(f)/f^2$.
\end{abstract}

\preprint{1901.09624}

\maketitle
\end{CJK*}

\section{Introduction}
The discovery of gravitational waves (GWs) by the Laser Interferometer Gravitational-Wave Observatory(LIGO) Scientific Collaboration and the Virgo Collaboration
started the new era of multimessenger astronomy and opened a new window to test
general relativity and probe the nature of gravity in the strong field regime \cite{Abbott:2016blz,Abbott:2016nmj,Abbott:2017vtc,Abbott:2017oio,TheLIGOScientific:2017qsa,Abbott:2017gyy,LIGOScientific:2018mvr}.
GWs propagate with the speed of light transverse to the direction of propagation and they have only two tensor polarizations
in the theory of general relativity.
Due to the possible existence of extra fields,
GWs in alternative theories beyond general relativity may have up to six polarizations \cite{Eardley:1974nw,Liang:2017ahj,Hou:2017bqj,Gong:2017bru,Gong:2018cgj,Gong:2018ybk,
Gong:2018vbo,Hou:2018djz,Hou:2018djz}, so the nature of gravity can be probed by the detection of the polarization content of GWs.
To detect the polarization content of short duration GWs and probe the nature of gravity, more detectors are needed.
The ground-based detectors, such as Advanced LIGO \cite{Harry:2010zz,TheLIGOScientific:2014jea}, Advanced Virgo \cite{TheVirgo:2014hva} and KAGRA \cite{Somiya:2011np,Aso:2013eba}, operating in the high-frequency band (10-$10^4$ Hz),
will form a network of detectors to probe the polarizations of GWs.
For continuous GW signals like those from binaries that are widely separated, one detector with changing orientation with respective to the source is enough.
The proposed space-based detectors such as LISA \cite{Danzmann:1997hm,Audley:2017drz}, TianQin \cite{Luo:2015ght} and TaiJi \cite{Hu:2017mde} probe GWs in the frequency band of millihertz, while DECIGO  \cite{Kawamura:2011zz} operates in the frequency band of 0.1 to 10 Hz.
The LIGO/Virgo collaborations found that
the pure tensor polarizations were favored against pure vector and pure scalar polarizations \cite{Abbott:2017oio,Abbott:2018lct}.

For ground-based detectors, the physical arm length of the detector is always shorter
than the wavelength of  in-band GWs (with frequencies up to $10^3$ Hz). When the wavelength of GWs becomes comparable
or even shorter than the physical arm length for the space-based detectors,
a deterioration of the signal-to-noise
ratio (SNR) registered in the detector occurs due to the decrease of the antenna transfer function at high frequencies.
It was pointed out that large systematic biases will be introduced if neglecting the frequency-dependent response for the third generation ground-based detectors \cite{Essick:2017wyl}.
Considering the case that the physical arm length is comparable to the wavelength of GWs,
the angular and frequency response of LISA in the ecliptic coordinate system was derived in \cite{Schilling:1997id,Giampieri:1997kv}.
In the wave coordinate system, a simple analytic formula for the averaged response function for the tensor mode was obtained
and the sensitivity curve for space-based detectors was discussed in \cite{Larson:1999we}.
For the discussion on the coordinate systems, please see Refs. \cite{Cutler:1997ta,Barack:2003fp}.
The sensitivity with various power spectra of noises to different polarizations
in different frequencies for LISA was studied in \cite{Tinto:2010hz}.
There are also discussions on the response functions of arbitrary polarizations for
other detectors satisfying the low frequency limit \cite{Nishizawa:2009bf,Hayama:2012au,Nishizawa:2009jh}.
In \cite{Blaut:2012zz}, the author analyzed the angular and
frequency behavior and the sensitivity patterns of the responses for arbitrary polarizations with different detector designs.
For more discussions on the construction and use of sensitivity curves for LISA, please see Ref. \cite{Cornish:2018dyw}.

Apart from compact binary coalescence, stochastic gravitational wave background is another important source for gravitational wave detectors.
Although the stochastic gravitational wave background has not been detected yet, it can be used to probe the Universe at very early times.
To characterize the stochastic gravitational wave background, the cross correlation
between two detectors and the complete analyses of the noises are needed \cite{Christensen:1992wi,Flanagan:1993ix,Allen:1997ad,Ungarelli:2000jp,Cornish:2001qi,Kudoh:2005as,Corbin:2005ny,
Finn:2008vh,Adams:2010vc,Adams:2013qma,Smith:2016jqs,Callister:2017ocg,Romano:2016dpx}.
For different sources, it is possible to measure the polarizations by combining a network of detectors \cite{Isi:2017fbj,Isi:2017equ,Takeda:2018uai,Hagihara:2018azu}.
The analyses of stochastic gravitational wave background with data from the two LIGO detectors show no evidence of any polarization
and the 95\% limits are $\Omega_0^T<5.58\times 10^{-8}$, $\Omega_0^V<6.35\times 10^{-8}$, and $\Omega_0^S<1.08\times 10^{-7}$
at a reference frequency $f_0=25$ Hz \cite{Abbott:2018utx}.

Although there are lots of discussions on the frequency dependence of the
antenna response functions for different polarizations \cite{Blaut:2012zz},
but there still lack similar analytic formulas for nontensorial polarizations.
In this paper, we extend the analysis of the frequency response for the tensor mode in \cite{Larson:1999we}
and derive the analytic formulas for the averaged antenna response functions for non-tensorial polarizations.
We also discuss different coordinate systems
and the independence of the results on different coordinate systems.
Due to the modulations of the observed signal by the orbital motion of LISA,
we can locate the sources of GWs \cite{Cutler:1997ta,Cornish:2002rt}. For TianQin, the current plan is for the detector plane to always point in the same direction,
the frequency and angular response for TianQin needs to be discussed in detail,
so we use the proposed TianQin as an example although the results of this paper are applicable to any equal-arm
interferometric GW detectors without optical cavities in the arms.

The paper is organized as follows. In Sec. IIA, we present the basic configurations of TianQin and LISA,
the polarization tensors are briefly discussed in Sec. IIB
and the basic property of the response function and its behavior at the low frequency limit in the detector frame are discussed in Sec. IIC.
The angular antenna responses for different polarizations for TianQin and LISA in the ecliptic frame are discussed in Sec. III. In Sec. IV,
the averaged response functions for different polarizations in the wave frame are derived and the sensitivity curves for TianQin are presented.
We conclude the paper in Sec. V. The orbits for LISA and TianQin are discussed in Appendix A, the coordinate systems and their relationships
are discussed in Appendix B.

\section{Background}

\subsection{The configurations of TianQin and LISA}

\begin{figure}[htp]
	\centering
	\includegraphics[width=0.5\textwidth]{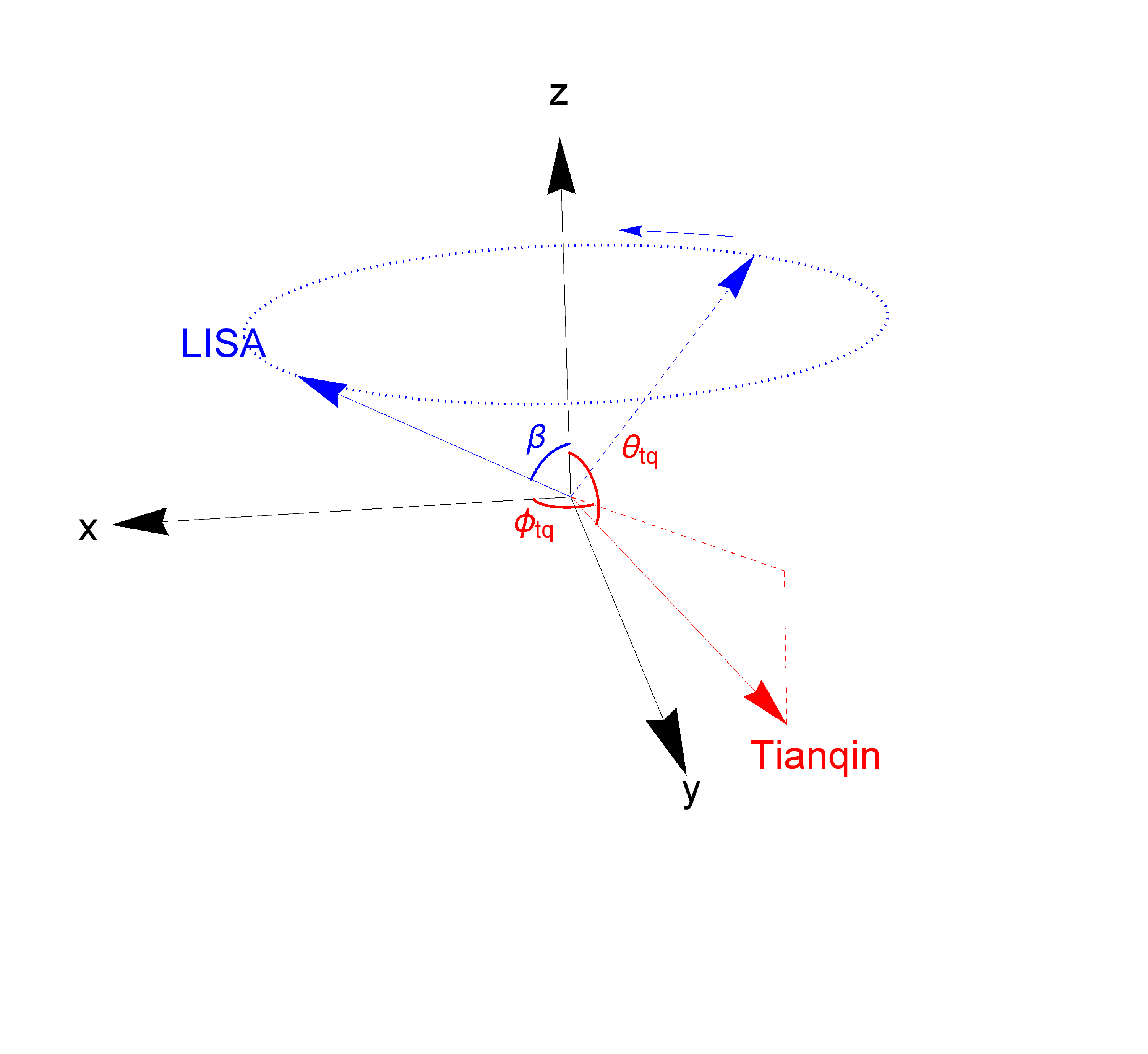}
\caption{The normal vectors of the detector planes for TianQin and LISA in ecliptic coordinates. TianQin points to
the direction of J0806.3+1527 located at ($\theta_{tq}=94.7^\circ$, $\phi_{tq}=120.5^\circ$).}
\label{TQLS}
\end{figure}

TianQin is formed by three identical drag-free spacecrafts in an equilateral triangular
constellation with sides of approximately $1.7\times 10^5$ km orbiting around the
Earth with a period of 3.65 days. The detector plane is fixed and initially designed toward the source J0806.3+1527. LISA's three spacecrafts orbit the
Sun in a nearly Earth-trailing orbit and their relative orbits define an equilateral triangle with sides of $2.5\times 10^6$ km and a period of 365.2422 days. The detector plane is inclined $60^\circ$ with respect to the ecliptic plane.
That means the normal vector of LISA's detector plane will span a circular cone in one year with an angle of $\beta=60^\circ$ between the cone surface and the $z$ axis of the ecliptic plane,
while the normal vector of TianQin's detector plane does not change with time as shown in Fig. \ref{TQLS}.

\subsection{The polarization tensors}
We use a set of unit vectors $\{\hat{p},\hat{q},\hat{w}\}$ which are perpendicular to each other
to form an orthonormal coordinate system such that $\hat{w}=-\hat{\Omega}$ is
the propagating direction of GWs and $\hat{w}=\hat{p}\times\hat{q}$.
Here $\hat{\Omega}$ is the direction of the source.
There is a rotational degree of freedom for the choices of $\hat{p}$ and $\hat{q}$
around $\hat{w}$ axis, which depends on the orientation of the source and is referred to as the polarization angle $\psi$,
\begin{equation}
\label{polarizationangle}
\hat{m}=\cos\psi \hat{p}+\sin \psi \hat{q}, \qquad \hat{n}=-\sin\psi \hat{p}+\cos\psi \hat{q}.
\end{equation}
The unit vectors $\hat{p}$ and $\hat{q}$ can be fixed by choosing a convention,
for the choice of a particular convention, please see Refs. \cite{Apostolatos:1994mx,Sathyaprakash:2009xs,Yunes:2013dva,Romano:2016dpx}.
Thus, we use $\{\hat{m},\hat{n},\hat{w}\}$ to define the six polarization tensors as follows
\begin{equation}
\begin{aligned}
e^{+}_{ij}=\hat{m}_i\hat{m}_j-\hat{n}_i\hat{n}_j,  &\qquad  e^{\times}_{ij}=\hat{m}_i\hat{n}_j+\hat{n}_i\hat{m}_j, \\
e^{x}_{ij}=\hat{m}_i\hat{w}_j+\hat{w}_i\hat{m}_j, &\qquad   e^{y}_{ij}=\hat{n}_i\hat{w}_j+\hat{w}_i\hat{n}_j, \\
e^{b}_{ij}=\hat{m}_i\hat{m}_j+\hat{n}_i\hat{n}_j, &\qquad  e^{l}_{ij}=\hat{w}_i\hat{w}_j.
\end{aligned}
\end{equation}

\subsection{The response function}

The signal for a single round-trip in the arm of the interferometric GW detector without optical cavities is defined as
\begin{equation}
s(t)=\sum_A F^A h_A(t),
\end{equation}
where $A=+,\times,x,y,b,l$ stands for the six  polarizations, $h_A(t)$ is the input signal of GWs,
and $F^A$ is the angular response function for polarization $A$
\begin{equation}
F^A=D^{ij} e^A_{ij}.
\end{equation}
The detector tensor $D^{ij}$ is given by
\begin{equation}
D^{ij}=\frac{1}{2}[\hat{u}^i \hat{u}^j T(f,\hat{u}\cdot\hat{w})-\hat{v}^i \hat{v}^j  T(f,\hat{v}\cdot\hat{w})],
\end{equation}
where $\hat{u}$ and $\hat{v}$ are the unit vectors along the arms of the detector and $T(f,\hat{u}\cdot\hat{w})$ is \cite{Cornish:2001qi,Estabrook:1975}
\begin{equation}
\label{transferfunction}
\begin{split}
T(f,\hat{u}\cdot\hat{w})=\frac{1}{2}&\left\{\text{sinc}\left[\frac{f}{2f^*}(1-\hat{u}\cdot\hat{w})\right]\exp\left[-i\frac{f}{2f^*}(3+\hat{u}\cdot\hat{w})\right] \right.\\
                            & +\left. \text{sinc}\left[\frac{f}{2f^*}(1+\hat{u}\cdot\hat{w})\right]\exp\left[-i\frac{f}{2f^*}(1+\hat{u}\cdot\hat{w})\right]\right\}.
\end{split}
\end{equation}
Here $\text{sinc}(x)=\sin x/x$, $f^*=c/(2\pi L)$ is the transfer frequency of the detector and $L$ is the arm length of the detector.
Note that in the low frequency limit $f\ll f^*$, we have $T(f,\hat{u}\cdot\hat{w})\rightarrow1$.
It is obvious that the angular response function is a complex scalar under the three-dimension coordinate rotation,
thus we can calculate it in any coordinate system. We also introduce the angular response function $F^t$
for the combined tensor mode
\begin{equation}
\label{fteq1}
F^t=\sqrt{|F^+|^2+|F^\times|^2},
\end{equation}
and the angular response function $F^v$ for the combined vector mode,
\begin{equation}
\label{fveq1}
F^v=\sqrt{|F^x|^2+|F^y|^2}.
\end{equation}
Since the detector tensor is symmetric, using Eq. \eqref{polarizationangle}, we obtain
\begin{equation}
|F^t|^2 =|\hat{p}_i \hat{p}_j D^{ij}-\hat{q}_i \hat{q}_j D^{ij}|^2+4|\hat{p}_i \hat{q}_j D^{ij}|^2,
\end{equation}
so the angular response function for the combined tensor mode is independent of the polarization angle $\psi$.
Similarly, one can show that the angular response functions for the combined vector, breathing and
longitudinal modes are also independent of the polarization angle $\psi$.
Furthermore, when averaging over the polarization angle $\psi$, it is easy to verify that
\begin{equation}
\label{averagetensor}
\langle|F^+|^2\rangle_{\psi} \equiv\frac{1}{2\pi} \int_0^{2\pi}|F^+|^2 d\psi =\langle|F^\times|^2\rangle_{\psi}=\frac{1}{2}\langle|F^t|^2\rangle_{\psi}=\frac{1}{2}|F^t|^2,
\end{equation}
and
\begin{equation}
\label{averagevector}
\langle|F^x|^2\rangle_{\psi}=\langle|F^y|^2\rangle_{\psi}=\frac{1}{2}|F^v|^2.
\end{equation}

In low frequency limit, $T\rightarrow 1$, we get the angular response functions for different polarizations in the detector frame
(the frequency dependent results are presented in Appendix \ref{coordsys}.)
\begin{equation}
\label{fplus2}
F^+=-\frac{\sin\gamma}{2}[(1+\cos^2\theta_d)\sin2\phi_d\cos2\psi+2\cos\theta_d\cos2\phi_d\sin2\psi],
\end{equation}
\begin{equation}
\label{fcross2}
F^\times=-\frac{\sin\gamma}{2}[-(1+\cos^2\theta_d)\sin2\phi_d\sin2\psi+2\cos\theta_d\cos2\phi_d\cos2\psi],
\end{equation}
\begin{equation}
\label{fvecx2}
F^x=-\frac{\sin\gamma}{2}[\sin2\theta_d\sin2\phi_d\cos\psi+2\sin\theta_d\cos2\phi_d\sin\psi],
\end{equation}
\begin{equation}
\label{fvecy2}
F^y=-\frac{\sin\gamma}{2}[-\sin2\theta_d\sin2\phi_d\sin\psi+2\sin\theta_d\cos2\phi_d\cos\psi],
\end{equation}
\begin{equation}
\label{Fl}
F^l=-F^b=-\frac{\sin\gamma}{2}\sin^2\theta_d\sin2\phi_d,
\end{equation}
where $\gamma$ is the angle between the two arms of the detector,
 $(\theta_d,\phi_d)$ are the coordinates of the location of the source in detector frame and  $\psi$ is the polarization angle.
If the detector arms are separated by $\gamma=\pi/2$, then we recover the angular response function for
second generation ground-based detectors such as Advanced LIGO.
It is clear that, up to a minus sign, the angular response functions for the breathing mode and the longitudinal mode are degenerate in the low frequency limit.
However, the response functions for the breathing and longitudinal
modes become totally different at high frequencies (see Appendix \ref{coordsys} for details).
In the low frequency limit, the angular response functions for the combined tensor and the combined vector modes are
\begin{equation}
\label{Ft}
|F^t|^2=\frac{\sin^2\gamma}{4}[(1-\cos^2\theta_d)^2\sin^22\phi_d+4\cos^2\theta_d],
\end{equation}
\begin{equation}
\label{Fv}
|F^v|^2=\sin^2\gamma\sin^2\theta_d(1-\sin^2\theta_d\sin^22\phi_d).
\end{equation}
Based on these results, we find the locations of the minima and maxima of the response functions.
In the detector frame, when $\theta_d=\{0, \pi\}$, we get $|F^t|_{max}=\sin\gamma$,
when $\phi_d=\{0,\pi/2,\pi,3\pi/2\}$  and $\theta_d=\pi/2$, we get $|F^t|_{min}=0$;
when $\phi_d=\{0,\pi/2,\pi,3\pi/2\}$ and $\theta_d=\pi/2$, we get $|F^v|_{max}=\sin\gamma$,
when $\theta_d=\{0, \pi\}$, or $\theta_d=\pi/2$ and $\phi_d=\{ \pi/4,3\pi/4,5\pi/4,7\pi/4\}$, we get $|F^v|_{min}=0$;
when $\theta_d=\pi/2$ and $\phi_d=\{\pi/4,3\pi/4,5\pi/4,7\pi/4\}$, we get $|F^b|_{max}=|F^l|_{max}=\sin\gamma/2$,
when  $\theta_d=\{0, \pi\}$ or $\phi_d=\{0,\pi/2,\pi,3\pi/2\}$ , we get  $|F^b|_{min}=|F^l|_{min}=0$.

\section{The angular antenna response}

\begin{figure}[htp]
\includegraphics[width=0.8\textwidth]{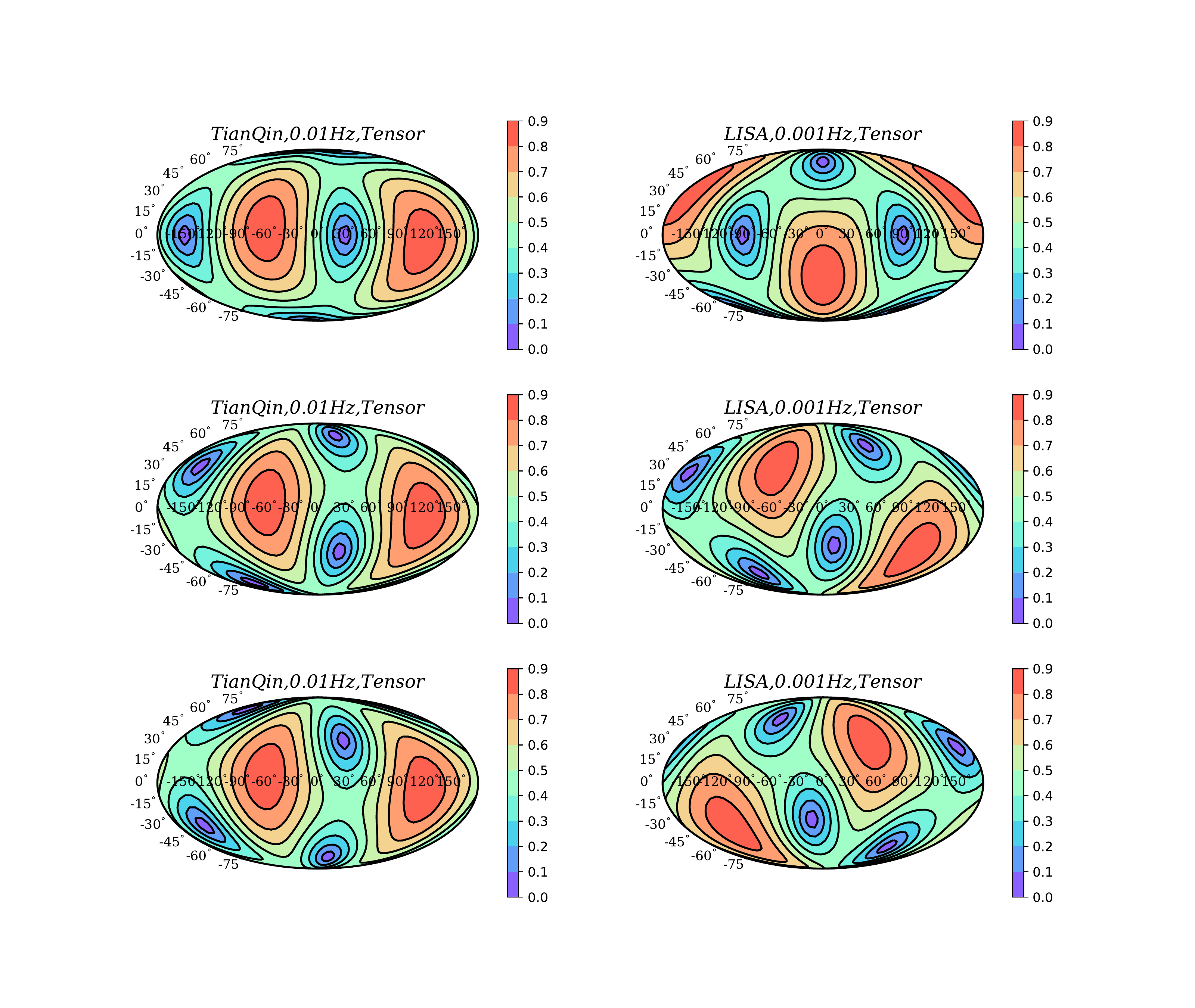}
\caption{The evolution of the angular response function for the combined tensor mode in ecliptic coordinates during the orbital motion of the detector.
The left panel is for TianQin and the right panel is LISA.
From the top to the bottom, the time interval is one third of the orbital period of the detector (1.22 day for TianQin and 4 months for LISA).
TianQin always points to the same direction, the most sensitive regions do not change.
For LISA, the most sensitive regions move along the latitude of $\pm 30^\circ$. For the animations of the plots, please see http://doi.org/10.5281/zenodo.2574620.}
\label{figtensor1}
\end{figure}

The critical frequency corresponding to the transfer frequency $f^*=c/(2\pi L)$ is 0.28Hz for TianQin,
and 0.019Hz for LISA. As long as the frequency of GWs is well below $f^*$,
the response function for interferometric GW detectors without optical cavities is independent of frequency as seen
from Eqs. \eqref{fplus2}-\eqref{Fl}.
When the wavelength of GWs is comparable with or shorter than the arm length L of the detector,
the antenna response function starts to decrease with the frequency which deteriorates the detector's performance.
In this section, we show the effect of the orbital motion of the detector on the antenna response function.
The detailed orbital motions are presented in Appendix \ref{orbits}.
We choose the frequency 0.01Hz for TianQin and 0.001Hz for LISA to compare their responses.
Following \cite{Romano:2016dpx}, in ecliptic coordinates we present the angular response function in Mollweide projection.
Three graphs with the time interval of one third of the orbital period of the detector (1.22 day for TianQin and 4 months for LISA) are shown
for each polarization mode.
In Fig. \ref{figtensor1}, we show the angular response for the combined tensor mode.
As explicitly shown in Eq. \eqref{Ft} and discussed in the last section,  in the low frequency limit,
the response function reaches the maximum (shown as the two red patches in Fig. \ref{figtensor1})
when the direction of the source is perpendicular or nearly perpendicular to the detector's plane.
Therefore, the location of the maximal response in the sky represents the normal direction of the detector's plane.
For TianQin, the normal vector of the detector's plane does not change will time,
so the maxima of the angular response function remain the same when the detector moves.
For LISA, as shown in Fig. \ref{figtensor1}, the most sensitive regions move along the latitude of $\pm 30^\circ$.
On the other hand, when GWs propagate along the detector's angular bisector
or the vertical direction of the angular bisector in the detector plane,
the response is the weakest (the 4 purple patches shown in Fig. \ref{figtensor1}).

\begin{figure}[htp]
	\includegraphics[width=0.8\textwidth]{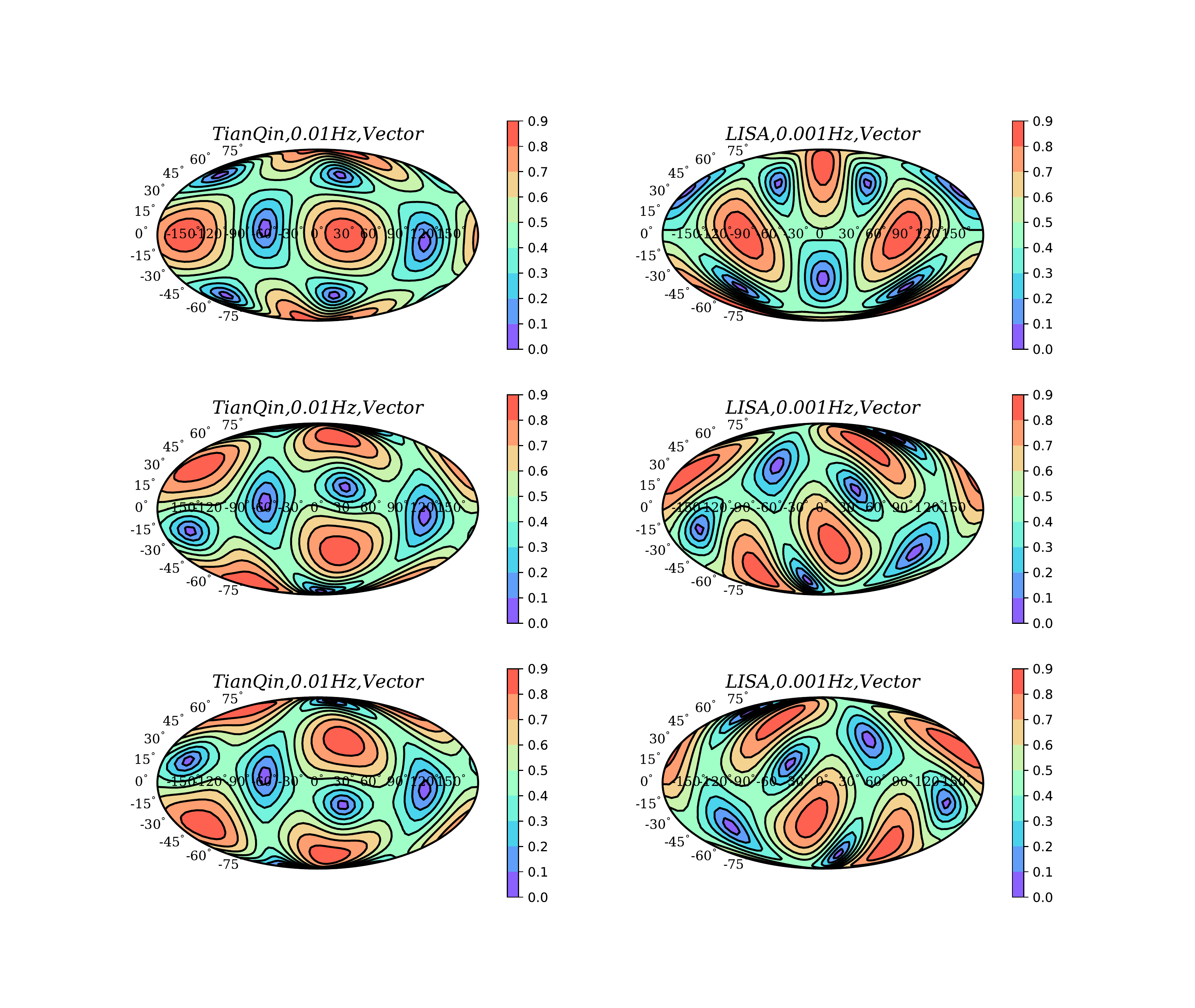}
\caption{The evolution of the angular response function for the combined vector mode in ecliptic coordinates during the orbital motion of the detector.
The left panel is for TianQin and the right panel is LISA.
From the top to the bottom, the time interval is one third of the orbital period of the detector (1.22 day for TianQin and 4 months for LISA).}
\label{figvector1}
\end{figure}

For the antenna response to (hypothetical) vector-polarized GWs, Eq. \eqref{Fv} tells us that the response function reaches the maximum
when GWs propagate along the detector's angular bisector or the vertical direction of the angular bisector in the detector plane,
and it becomes zero if GWs propagate along the direction perpendicular to the detector's plane, which is exactly opposite to the tensor mode. There are four more insensitive regions along the direction of the detector plane, which are located between the sensitive regions.
In Fig. \ref{figvector1}, we show the angular response for the combined vector mode.

\begin{figure}[htp]
	\includegraphics[width=0.8\textwidth]{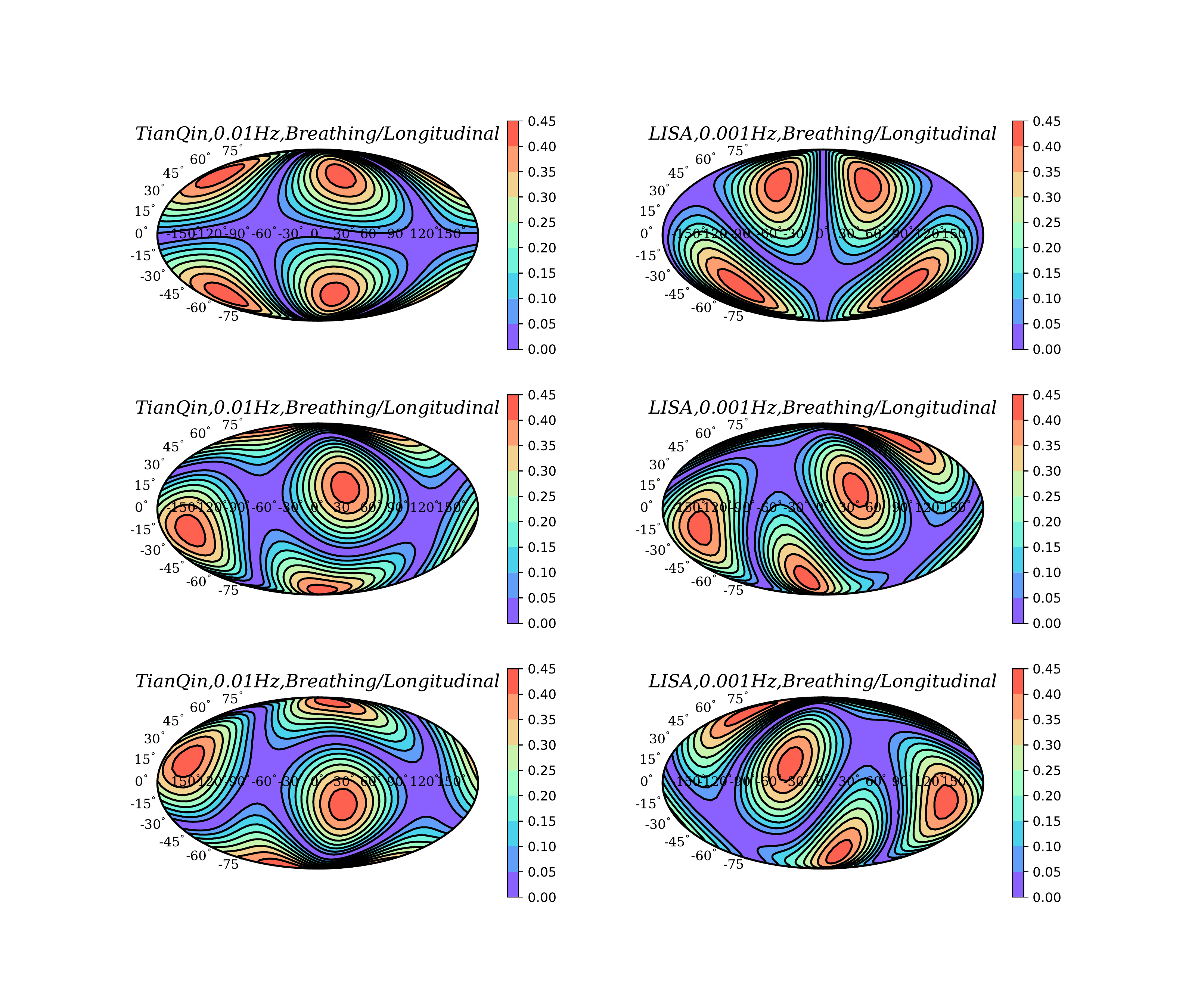}
\caption{The evolution of the angular response function for the breathing/longitudinal mode in ecliptic coordinates during the orbital motion of the detector.
The left panel is for TianQin and the right panel is LISA.
From the top to the bottom, the time interval is one third of the orbital period of the detector (1.22 day for TianQin and 4 months for LISA).}
\label{figscalar1}
\end{figure}

From Eq. \eqref{Fl}, we see that the four most sensitive regions for the breathing/longitudinal
mode are exactly the insensitive regions for the vector mode.
We show the angular response for the breathing/longitudinal mode in Fig. \ref{figscalar1}.

\begin{figure}[htp]
	\includegraphics[width=0.8\textwidth]{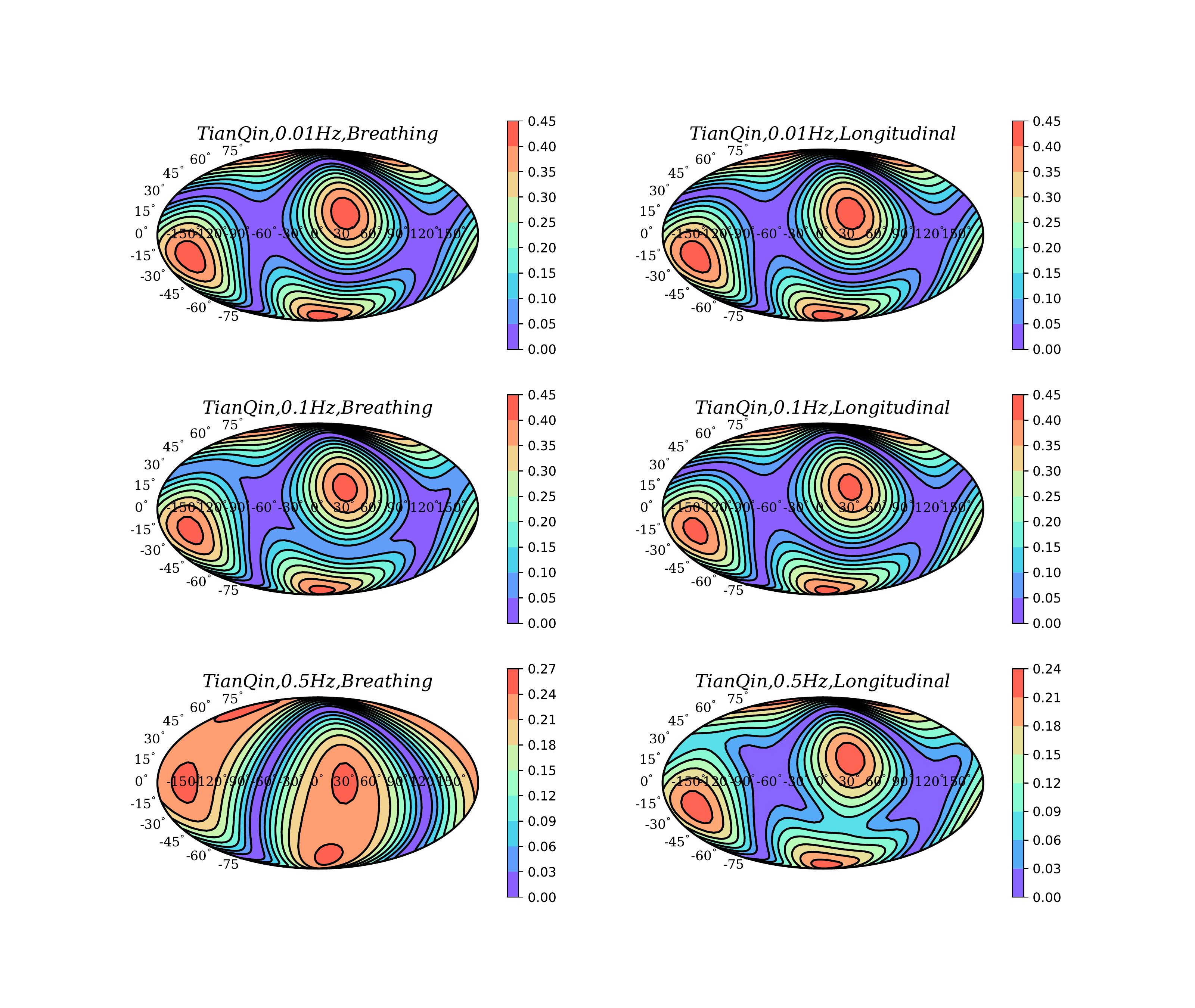}
\caption{The frequency dependence of the TianQin's response function for the breathing and longitudinal modes.
The left panel is for the breathing mode and the right panel is for the longitudinal mode. From the top to the bottom,
three different frequencies of 0.01Hz, 0.1Hz and 0.5Hz are chosen, respectively.}
\label{figscalar2}
\end{figure}

Finally, we compare the frequency dependence of the TianQin's response function for the breathing and longitudinal modes.
Three different frequencies of 0.01Hz, 0.1Hz and 0.5Hz are chosen. The results are shown in Fig. \ref{figscalar2}.
It is obvious that the degeneracy between the breathing and longitudinal modes are broken at high frequencies.

\section{The averaged response function and the sensitivity curves}

Generally, we do not know where GWs come from, so we average over the source locations and the polarization angles
to investigate how the detector responds to different polarizations in different frequencies and obtain the respective sensitivity,
\begin{equation}
R_A(f)=\frac{1}{8\pi^2}\int\int |F^A|^2 d\Omega d\psi.
\end{equation}
The averaged response function $R_A(f)$ relates the power spectral density of the incident gravitational wave signals to the power spectral density of the signal recorded in the detector.

In the ecliptic coordinate system, the averaged response function is given by
\begin{equation}
\label{raverage6}
R_A(f)=\frac{1}{8\pi^2}\int_0^{2\pi} \int_0^{2\pi}\int_0^\pi |F^A|^2 \sin\theta_e d\theta_e d\phi_e d\psi_e,
\end{equation}
where $A=+,\times,x,y$, ($\theta_e$, $\phi_e$) and $\psi_e$ are the direction and the polarization angle of the source, respectively.
As discussed in the last section, the tensor, vector, breathing and longitudinal modes are independent of the polarization angle $\psi_e$,
so we average the source direction $(\theta_e,\phi_e)$ only,
\begin{equation}
\label{raverage7}
R_A(f)=\frac{1}{4\pi}\int_0^{2\pi} \int_0^\pi |F^A|^2 \sin\theta_e d\theta_e d\phi_e,
\end{equation}
where $A=t,v,b,l$.
Using Eqs. \eqref{averagetensor} and \eqref{averagevector}, we get
\begin{equation}
\label{raverage8}
R_+(f)=R_\times(f)=\frac{1}{2}R_t(f), \qquad R_x(f)=R_y(f)=\frac{1}{2}R_v(f).
\end{equation}
When doing the numerical integration, it costs less time to do the double integral than to do the triple integral.
Thus, instead of calculating $R_+$, $R_\times$, $R_x$ or $R_y$ using Eq. \eqref{raverage6},
we use Eq. \eqref{raverage7} to calculate $R_t$ or $R_v$ first, and then obtain $R_+$,
$R_\times$, $R_x$ or $R_y$ from Eq. \eqref{raverage8}.

In the detector coordinate system, the averaged angular response function is given by
\begin{equation}
\label{raverage9}
R_A(f)=\frac{1}{4\pi} \int_0^{2\pi}\int_0^\pi |F^A|^2 \sin\theta_d d\theta_d d\phi_d,
\end{equation}
where $A=t,v,b,l$.
In the low frequency limit, inserting Eqs. (\ref{Fl}-\ref{Fv}) into the above Eq. \eqref{raverage9}, we get
\begin{equation}
\label{Rp}
R_+=R_\times=R_x=R_y=\frac{1}{5}\sin^2\gamma,
\end{equation}
and
\begin{equation}
\label{Rb}
R_b=R_l=\frac{1}{15}\sin^2\gamma.
\end{equation}
Since in the detector coordinate system, the detector is fixed while the source is moving,
it is clear that after averaging over all source locations, the angular response function
is independent of time if the arm length and the opening angle $\gamma$ between the two arm keep constant with time.
The detailed discussions on the coordinate systems are presented in Appendix \ref{coordsys}.

\begin{figure}[htp]
\centering
\includegraphics[width=0.6\textwidth]{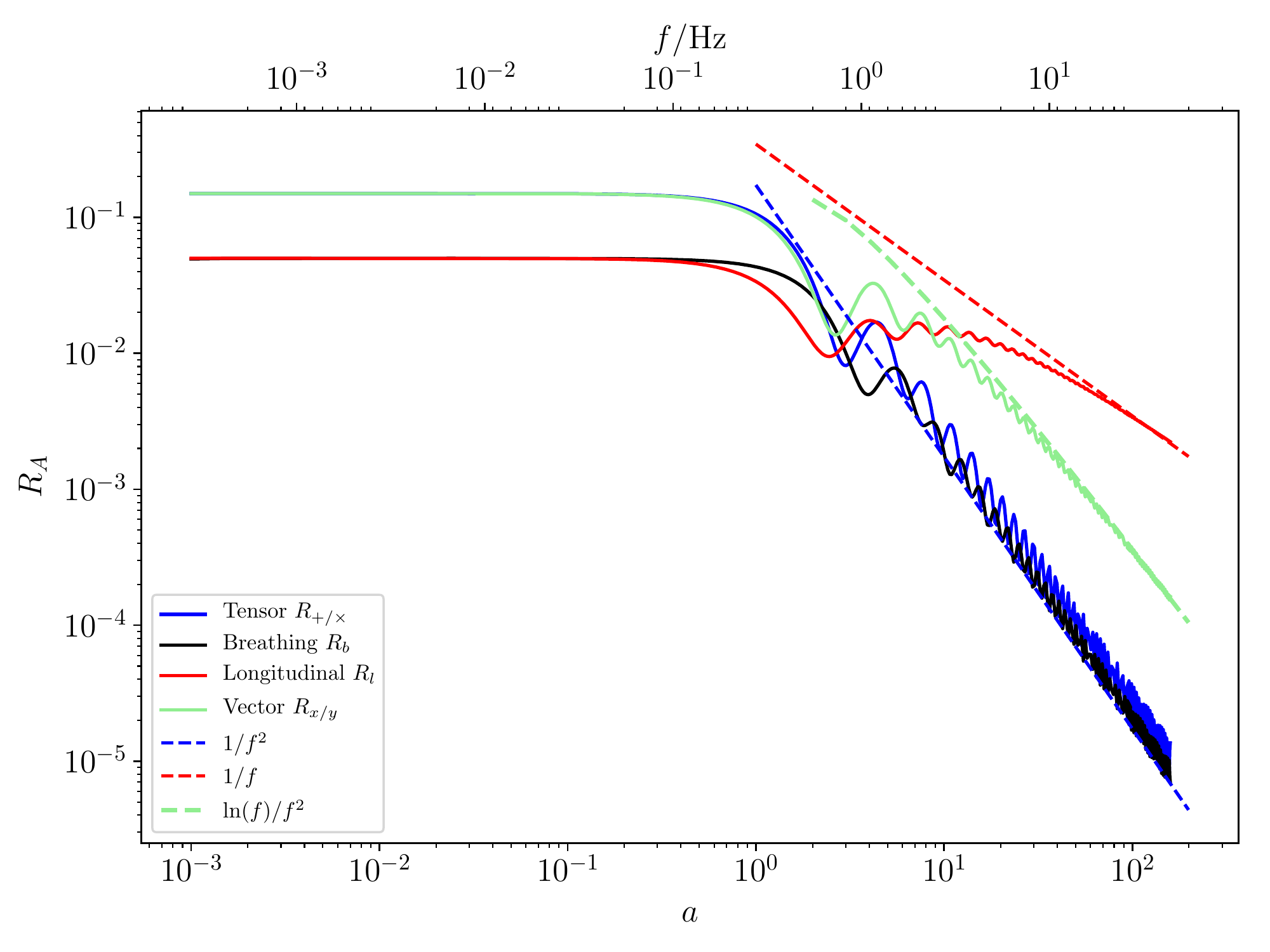}
\caption{The averaged antenna response functions for different polarizations.
 In the lower $x$ axis, we use the variable $a=2\pi f L/c$ and choose the opening angle $\gamma=\pi/3$.
The dashed curves show the asymptotic behaviors at high frequencies.
Similar results were obtained in \cite{Blaut:2012zz} by performing a Monte Carlo simulation,
here we use the analytical expressions (\ref{rteq6}-\ref{rleq6}).
 In the upper $x$ axis, we show TianQin's frequency dependent antenna response function for different polarizations.
For TianQin, the arm length is $L=\sqrt{3}\times 10^8$ m and the opening angle $\gamma=\pi/3$}.
\label{rufig1}
\end{figure}

In the wave coordinate system, the averaged angular response function is given by
\begin{equation}
\label{averagedplus}
R_A(f)=\frac{1}{8\pi^2}\int_0^{2\pi} \int_0^{2\pi}\int_0^\pi |F^A|^2 \sin\theta_1 d\theta_1 d\epsilon d\psi_1,
\end{equation}
where $A=+,\times,x,y,b,l$, and the angles $\theta_1$, $\psi_1$ and $\epsilon$ are defined in appendix \ref{coordsys}.
Following \cite{Larson:1999we}, substituting Eqs. (\ref{fplus0}-\ref{flong0}) to Eq. \eqref{averagedplus},
we derive the expressions of the averaged response functions for all six polarizations as
\begin{equation}
\label{rteq6}
\begin{split}
R_+(f)=&R_\times(f)=\frac{1}{2}R_t(f)\\
&=\frac{1}{4a^2}\left[ [1+\cos^2(a)] \left(\frac{1}{3}-\frac{2}{a^2} \right)\right.
+\frac{2}{a^3} \sin(2a) +\sin^2(a)\\
&-\left.\frac{1}{4\pi}\int_0^{2\pi}d\epsilon \int_0^\pi d\theta_1 \sin\theta_1 \left(1-\frac{2\sin^2\gamma\sin^2\epsilon}{1-\mu_2^2}\right)\eta(a)\right],
\end{split}
\end{equation}
\begin{equation}
\label{rveq6}
\begin{aligned}
R_x(f)&=R_y(f)=\frac{1}{2}R_v(f)\\
&=\frac{1}{2a^2}\left[-5+2\ln2+2\gamma_E+2\ln(a)-\frac{1}{3}\cos(2a)-2\text{Ci}(2a) \right.\\
 &+\frac{-4\sin(2a)+4[1+\cos^2(a)]a+2\sin(2a)a^2}{a^3} \\
 &\left.-\frac{1}{2\pi}\int_0^\pi d\theta_1 \int_0^{2\pi} d\epsilon
 \frac{\cos\theta_1(-\sin\gamma\cos\theta_1\cos\epsilon+\cos\gamma\sin\theta_1)\mu_2}{1-
 \mu_2^2} \eta(a) \right],
\end{aligned}
\end{equation}

\begin{equation}
\label{rbeq6}
\begin{split}
R_b(f)=\frac{1}{2a^2}&\left[ [1+\cos^2(a)] \left(\frac{1}{3}-\frac{2}{a^2} \right)\right. +\sin^2(a)\\
&+\frac{2}{a^3} \sin(2a)
-\left.\frac{1}{4\pi}\int_0^{2\pi}d\epsilon \int_0^\pi d\theta_1 \sin\theta_1 \eta(a)\right],
\end{split}
\end{equation}

\begin{equation}
\label{rleq6}
\begin{split}
R_l(f)=&\frac{1}{8a^2}\left[ 15-9\ln2-9\gamma_E-9\ln(a)+ \left(\frac{11}{3}-\ln 2-\gamma_E-\ln(a)\right)\cos(2a) \right.\\
&+[9+\cos(2a)]\text{Ci}(2a)+[2a+\sin(2a)]\text{Si}(2a) \\
&+\frac{8[\sin(2a)-(1+\cos^2(a))a-\sin(2a)a^2]}{a^3}\\
&-\left.\frac{1}{\pi}\int_0^\pi d\theta_1 \int_0^{2\pi} d\epsilon  \frac{\cos^2\theta_1\mu_2^2}{\sin\theta_1(1-\mu_2^2)}\eta(a) \right],
\end{split}
\end{equation}
where
\begin{equation}
\begin{aligned}
\eta(a)&=[\cos(a)-\cos(a\mu_1)][\cos(a)-\cos(a\mu_2)]\mu_1\mu_2 \\
&+[\sin(a)-\mu_1\sin(a\mu_1)][\sin(a)-\mu_2\sin(a\mu_2)],
\end{aligned}
\end{equation}
$\mu_1=\cos\theta_1$, $\mu_2=\cos\gamma\cos\theta_1+\sin\gamma\sin\theta_1\cos\epsilon$,
$\gamma_E$ is the Euler constant, $a=2\pi f L/c=f/f^*$, $\text{Si}(a)$
is sine-integral function and $\text{Ci}(a)$ is cosine-integral function.
In the low frequency limit $a\ll 1$, $\eta(a)\approx a^2(1-\mu_1^2)(1-\mu_2^2)$ and we recover the standard results \eqref{Rp} and \eqref{Rb}.
At high frequencies, the response functions for the tensor and breathing modes
fall off as $1/f^2$ as seen from Eqs. \eqref{rteq6} and \eqref{rbeq6},
the response functions for the vector and longitudinal modes fall off slower than those for the tensor and breathing modes
due to the $\ln(a)$ term in Eq. \eqref{rveq6} and the $a\text{Si}(a)$ term in Eq. \eqref{rleq6}.
The response function $R_l(f)$ for the longitudinal mode falls off as $1/f$ and the response function $R_v(f)$
for the vector mode falls off as $\ln(f)/f^2$.
We show the averaged angular response
function for different polarizations in Fig. \ref{rufig1}.
From Fig. \ref{rufig1}, we see that $R_A(a)$ is a constant at low frequencies and starts to
decrease at around $a=1$ or $f=f^*$,
the fall off behaviours at high frequencies confirm the above analyses and
these results are consistent with Eqs. \eqref{Rp}, \eqref{Rb} and (\ref{rteq6}-\ref{rleq6}).
Note that the results for the response
functions apply to any equal-arm interferometric gravitational-wave detectors without optical cavities in the arms.
In particular, we also show the frequency dependent antenna response function for TianQin in Fig. \ref{rufig1}.

\begin{figure}[htp]
	\centering
	\includegraphics[width=0.6\textwidth]{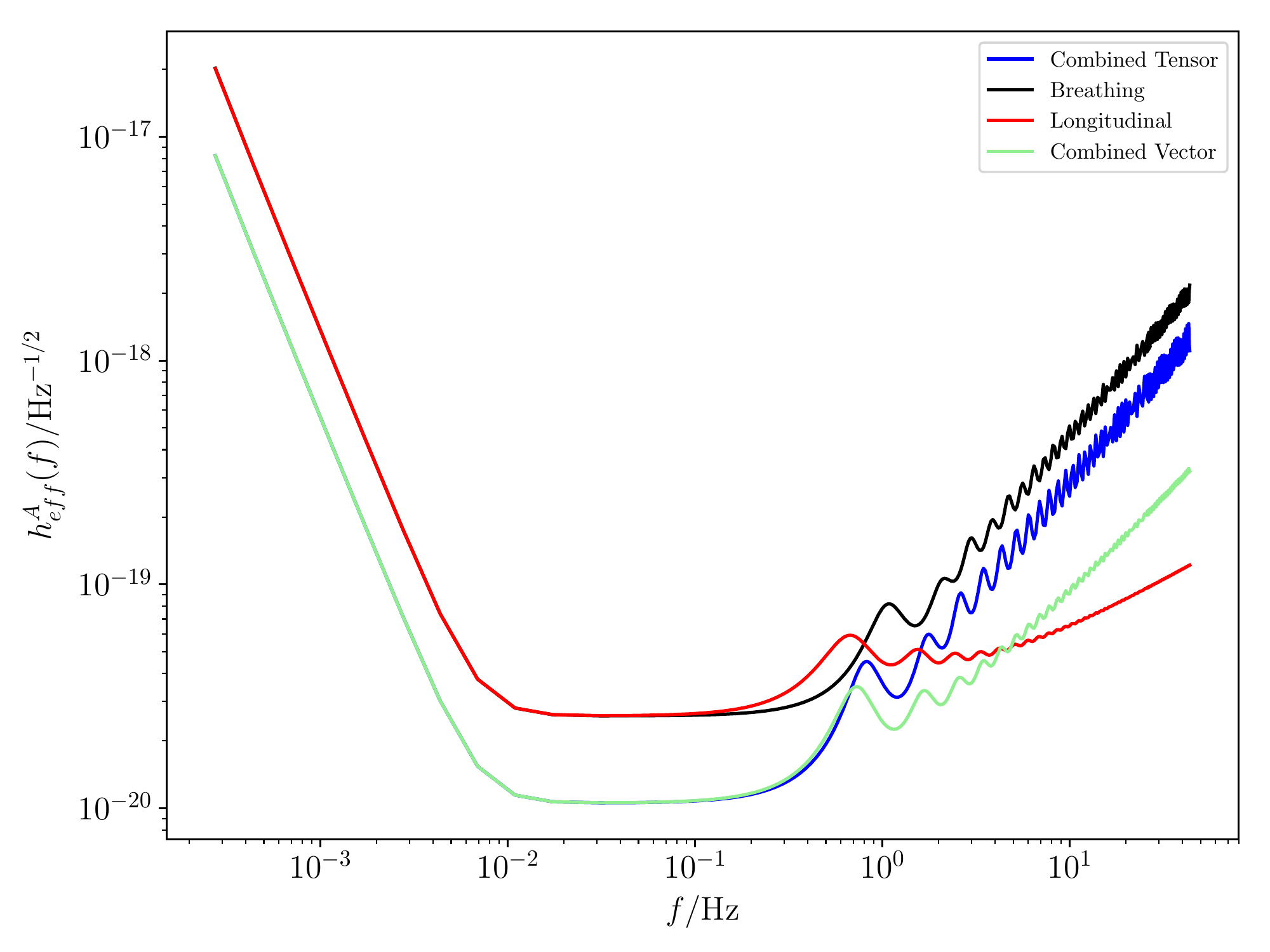}
\caption{The effective strain sensitivity curves for TianQin.}
\label{tqsenfig1}
\end{figure}

\begin{figure}[htp]
	\centering
	\includegraphics[width=0.6\textwidth]{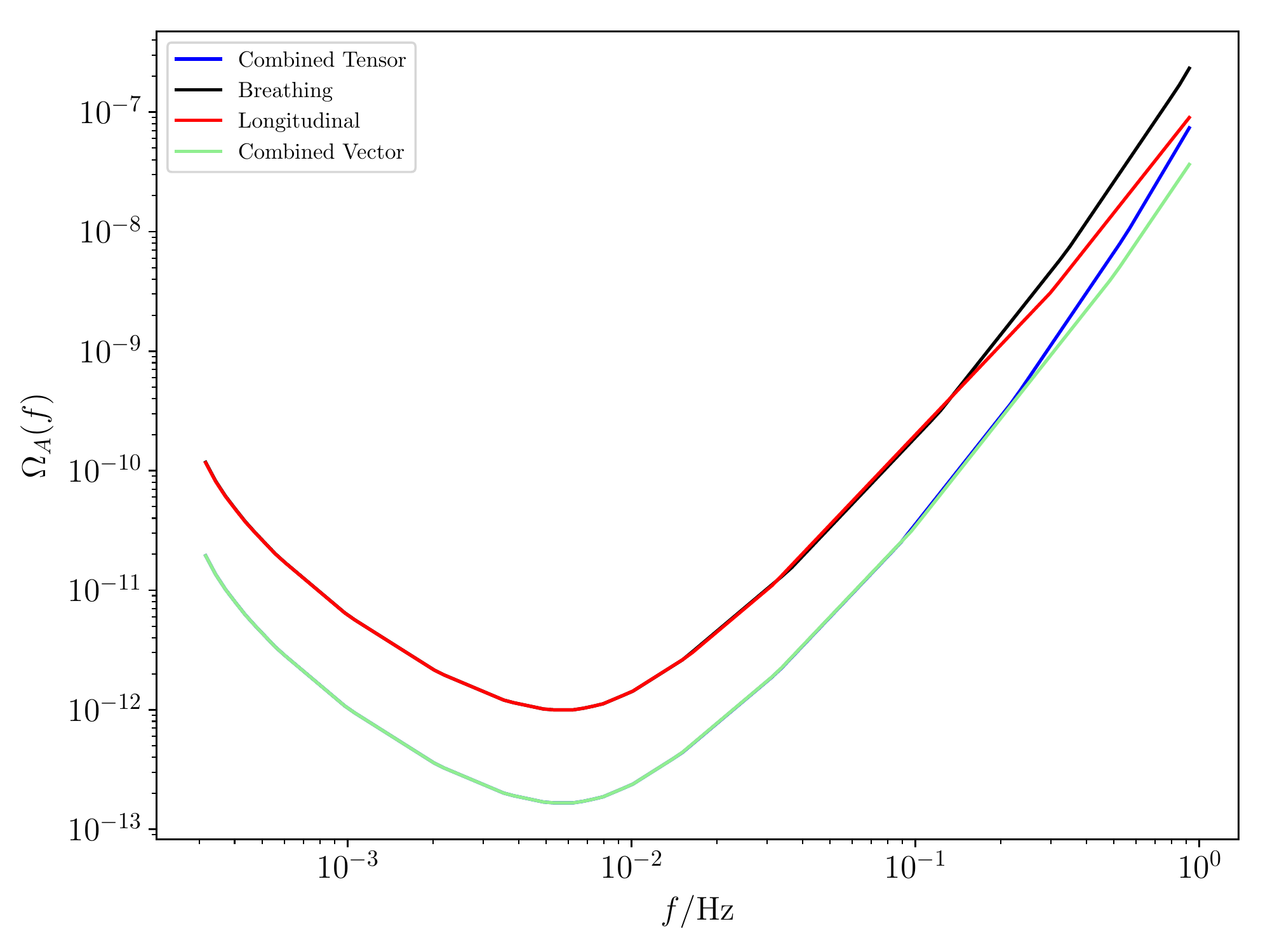}
\caption{The  power-law integrated curves characterizing the sensitivity of TianQin to a stochastic background for different polarizations.We choose SNR=1 and the observational time of one year.}
\label{tqsenfig2}
\end{figure}

Although the sensitivity curves are not useful for actual data analysis, they are usually
used to make a quick estimation of the detectability of the signal. Define the effective strain noise
\begin{equation}
h_{eff}^A(f)\equiv\sqrt{\frac{P_n(f)}{R_A(f)}},
\end{equation}
where the noise spectral density is \cite{Cornish:2001qi}
\begin{equation}
P_n(f)=\frac{S_x}{L^2}+\frac{4S_a}{(2\pi f)^4L^2}\left(1+\frac{10^{-4}\text{Hz}}{f}\right),
\end{equation}
the position noise $\sqrt{S_x}=1\text{pm/Hz}^{1/2}$ and the acceleration noise $\sqrt{S_a}=10^{-15}$m\,s$^{-2}$/Hz$^{1/2}$
for TianQin \cite{Luo:2015ght}, we plot the effective stain sensitivity curves for TianQin in Fig. \ref{tqsenfig1}.
Following \cite{Thrane:2013oya}, by taking SNR=1 and the observational time of one year,
we construct the power-law integrated sensitivity curves, to characterize the sensitivity to a stochastic GW background for TianQin in Fig. \ref{tqsenfig2}.

\section{Conclusion}
We describe the angular and frequency-dependent response function of
proposed space-based interferometric GW detectors LISA and TianQin,  for different GW polarizations. We present analytical formulas and sensitivity curves as functions of sky location and GW frequency. These functions are presented in terms of angles in three common coordinate systems.
The rotational matrices that transform one coordinate system to another one are also given explicitly.
Since the angular response function is a complex  scalar under three dimensional rotational transformation,
we can calculate it in any convenient coordinate system.
In ecliptic coordinates, we show how the most sensitive regions for different polarizations move with time for Tianqin and LISA.
For Tianqin, where it is assumed that the detector is always pointing at a single direction, at low frequencies the most sensitive regions for the tensor mode are independent of the orbital motions, so it maximizes the signal-to-noise if continuous GWs come from these regions.

In the detector frame, the detector is fixed while the source is moving with time relative to the detector.
Thus, the antenna response function averaging over the source directions is independent of the orbital motions of the detectors and time.
This conclusion is also independent of the choice of the coordinate systems.
We derive the analytical formulas for the frequency dependent response functions for all six polarizations which are valid for any
equal-arm interferometric GW detectors without optical cavities in the arms.
The response functions for the longitudinal and breathing modes are degenerate at low frequencies,
but they behave very differently at high frequencies. The analytical results show that
at high frequencies, the response functions for the tensor and breathing modes fall of as $1/f^2$,
the response function for the longitudinal mode falls off as $1/f$ and
the response function for the vector mode falls off as $\ln(f)/f^2$.
By using the derived response functions, we plot the frequency dependent response functions and the sensitivity curves for TianQin detector.
We also give the power-law integrated curves for TianQin to display its ability of detecting stochastic gravitational wave background.

\begin{acknowledgments}
This research was supported in part by the National Natural Science
Foundation of China under Grant No. 11875136,
the Major Program of the National Natural Science Foundation of China under Grant No. 11690021.
 A.J.W. acknowledges funding from the United States National Science Foundation, under Cooperative Agreement No. PHY-1764464.
\end{acknowledgments}

\appendix

\section{THE ORBITS OF LISA AND TIANQIN}
\label{orbits}
\subsection{TianQin's orbit}

In the ecliptic coordinate system, the TianQin's orbit $\vec{r}^T_n(t)=(X^{T}_n(t),Y^{T}_n(t),Z^{T}_n(t))$, ($n=1,2,3$) is \cite{Hu:2018yqb}
\begin{equation}
\begin{aligned}
X^{T}_n(t)=&R_T[\cos\theta_{tq}\cos\phi_{tq}\sin(\alpha_{Tn}-\alpha_{T0}')+\sin\phi_{tq}\cos(\alpha_{Tn}-\alpha_{T0}')] \\
&+\frac 12 e_TR_T[\sin\phi_{tq}(\cos 2(\alpha_{Tn}-\alpha_{T0}')-3)+\cos\theta_{tq}\cos\phi_{tq}\sin 2(\alpha_{Tn}-\alpha_{T0}')]\\
&+\frac 14e^2_TR_T[-6\sin\phi_{tq}\cos(\alpha_{Tn}-\alpha_{T0}') \sin^2(\alpha_{Tn}-\alpha_{T0}')\\
&+\cos\theta_{tq}\cos\phi_{tq}(3\cos 2(\alpha_{Tn}-\alpha_{T0}')-1)]\\
&+R\cos(\alpha_T-\alpha_{T0})+\frac 12 eR(\cos 2(\alpha_T-\alpha_{T0})-3)-\\
&\frac 32 e^2R\cos(\alpha_T-\alpha_{T0})\sin^2(\alpha_T-\alpha_{T0})+O(e^3,e^3_1),
\end{aligned}
\end{equation}

\begin{equation}
\begin{aligned}
Y^{T}_n(t)=&R_T[\cos\theta_{tq}\sin\phi_{tq}\sin(\alpha_{Tn}-\alpha_{T0}')-\cos\phi_{tq}\cos(\alpha_{Tn}-\alpha_{T0}')] \\
&+\frac 12 e_TR_T[-\cos\phi_{tq}(\cos 2(\alpha_{Tn}-\alpha_{T0}')-3)+\cos\theta_{tq}\sin\phi_{tq}\sin 2(\alpha_{Tn}-\alpha_{T0}')]\\
&+\frac 14e^2_TR_T[6\cos\phi_{tq}\cos(\alpha_{Tn}-\alpha_{T0}')\sin^2(\alpha_{Tn}-\alpha_{T0}')+\\
&\cos\theta_{tq}\sin\phi_{tq}\sin(\alpha_{Tn}-\alpha_{T0}')(3\cos 2(\alpha_{Tn}-\alpha_{T0}')-1)] \\
&+R\sin(\alpha_T-\alpha_{T0})+\frac 12 eR\sin 2(\alpha_T-\alpha_{T0})+\\
&\frac 14 e^2R\sin(\alpha_T-\alpha_{T0})(e\cos 2(\alpha_T-\alpha_{T0})-1)+O(e^3,e^3_1),
\end{aligned}
\end{equation}

\begin{equation}
\begin{aligned}
Z^{T}_n(t)=&-R_T\sin\theta_{tq}\sin(\alpha_{Tn}-\alpha_{T0}')-\frac 12 e_TR_T\sin\theta_{tq}\sin 2(\alpha_{Tn}-\alpha_{T0}')\\
&-\frac 14e^2_TR_T\sin\theta_{tq}\sin(\alpha_{Tn}-\alpha_{T0}')(3\cos 2(\alpha_{Tn}-\alpha_{T0}')-1),
\end{aligned}
\end{equation}
where ($\theta_{tq}=94.7^\circ$, $\phi_{tq}=120.5^\circ$) is the location of the source J0806.3+1527 for TianQin,
$e_T$ is the eccentricity of Tianqin's orbit which we set to be 0 in this paper,
$e=0.0167$ is the eccentricity of the Earth's orbit, $R_T=1.0\times 10^5$ km is
the semimajor axis of the spacecraft's orbit, $R=1$ AU,
$\alpha_{T0}$ and $\alpha'_{T0}$ are initial phases, $\alpha_T=2\pi f_mt+\kappa_0$, $f_m=1/(365$ days),
$\alpha_{Tn}=2\pi f_{sc}t+2\pi(n-1)/3$ and $f_{sc}=1/(3.65$ days).

\subsection{LISA's orbit}

The LISA's orbit $\vec{r}^L_n(t)=(X^{L}_n(t),Y^{L}_n(t),Z^{L}_n(t))$ in the ecliptic coordinate system is \cite{Dhurandhar:2004rv}
\begin{gather}
X^{L}_n(t)=R(\cos\alpha_{Ln}+e_L)\cos\epsilon\cos\theta_{Ln}-R\sqrt{1-e_L^2}\sin\alpha_{Ln}\sin\theta_{Ln},\\
Y^{L}_n(t)=R(\cos\alpha_{Ln}+e_L)\cos\epsilon\sin\theta_{Ln}+R\sqrt{1-e_L^2}\sin\alpha_{Ln}\cos\theta_{Ln},\\
Z^{L}_n(t)=R(\cos\alpha_{Ln}+e_L)\sin\epsilon,
\end{gather}
where $e_L=(1+\frac{2}{\sqrt{3}}\sigma+\frac{4}{3}\sigma^2)^\frac{1}{2}-1$,
$\epsilon=\arctan\frac{\sigma}{1+\sigma/\sqrt{3}}$,
$\sigma=\frac{L}{2R}$, $L=2.5\times10^9$ m is the arm length, $R\approx 1$ AU,
$\theta_{Ln}=2\pi(n-1)/3$, $\alpha_{Ln}+e\sin\alpha_{Ln}=2\pi f_m t-2\pi(n-1)/3-\alpha_{L0}$
and $\alpha_{L0}$ is the initial phase.

\section{THE COORDINATE SYSTEMS}

We use the index $e$ to denote quantities in the ecliptic coordinate system, $w$ for quantities in the wave
coordinate system, $d$ for quantities in the detector coordinate system,
 $c$ denotes the coordinate system in which the unit vectors $\hat{m}=(1,0,0)$, $\hat{n}=(0,1,0)$ and $\hat{w}=(0,0,1)$.
For the specific detectors, we use $t$ for quantities in TianQin's detector coordinates,
and $l$ for quantities in LISA's detector coordinates. The rotation matrix $R^{de}$ denotes the transformation from detector coordinates to ecliptic coordinates.

\label{coordsys}
\subsection{The ecliptic coordinate system}

In the ecliptic coordinate system, TianQin orbits the Earth, and LISA orbits the Sun on the ecliptic plane.
The unit vectors of the two arms of the detector in ecliptic coordinates $\{x_e,y_e,z_e\}$ are
\begin{gather}
\hat{u}_e(t)=\frac{\vec{r}_2(t)-\vec{r}_1(t)}{|\vec{r}_2(t)-\vec{r}_1(t)|}, \\
\hat{v}_e(t)=\frac{\vec{r}_3(t)-\vec{r}_1(t)}{|\vec{r}_3(t)-\vec{r}_1(t)|},
\end{gather}
where $\vec{r}_n(t)=\vec{r}^T_n(t)$ or $\vec{r}^L_n(t)$.
We label the location of an arbitrary source in ecliptic coordinates by the spherical coordinates $(\theta_e,\phi_e)$.
The unit vectors $\{\hat{m},\hat{n},\hat{w}\}$ describing the GWs are related to the ecliptic coordinates $\{x_e,y_e,z_e\}$ by the Euler rotation matrix
\begin{equation}
\begin{aligned}
R_{ij}^{ce}=
  \left(
    \begin{array}{cccc}
   \cos\theta_e\cos\phi_e\cos\psi_e+\sin\phi_e\sin\psi_e & -\cos\theta_e\cos\phi_e\sin\psi_e+\sin\phi_e\cos\psi_e &  -\sin\theta_e\cos\phi_e    \\
   \cos\theta_e\sin\phi_e\cos\psi_e-\cos\phi_e\sin\psi_e & -\cos\theta_e\sin\phi_e\sin\psi_e-\cos\phi_e\cos\psi_e & -\sin\theta_e\sin\phi_e \\
       -\sin\theta_e\cos\psi_e & \sin\theta_e\sin\psi_e & -\cos\theta_e     \\
    \end{array}
  \right),
\end{aligned}
\end{equation}
where $\psi_e$ is the polarization angle in ecliptic coordinates.
Using the rotation matrix, we express
the unit vectors describing the GWs in terms of the location of the source and the polarization angle
in ecliptic coordinates as
\begin{gather}
\begin{split}
\hat{m}=R^{ce}\cdot (1,0,0)=&
(\cos\theta_e\cos\phi_e\cos\psi_e+\sin\phi_e\sin\psi_e,\\
&\cos\theta_e\sin\phi_e\cos\psi_e-\cos\phi_e\sin\psi_e,-\sin\theta_e\cos\psi_e),
\end{split}\\
\begin{split}
\hat{n}=R^{ce}\cdot (0,1,0)
=&(-\cos\theta_e\cos\phi_e\sin\psi_e+\sin\phi_e\cos\psi_e,\\
&-\cos\theta_e\sin\phi_e\sin\psi_e-\cos\phi_e\cos\psi_e,\sin\theta_e\sin\psi_e),
\end{split}\\
\hat{w}=R^{ce}\cdot  (0,0,1)=(-\sin\theta_e\cos\phi_e,-\sin\theta_e\sin\phi_e,-\cos\theta_e).
\end{gather}
Plugging in the orbit coordinates,
the angular response functions for Tianqin and LISA for different polarizations are
\begin{equation}
\label{FAecliptic}
F^A=\frac{1}{2}\left[\hat{u}^i_e\hat{u}^j_e T(f,\hat{u}_e\cdot\hat{w}_e)- \hat{v}^i_e\hat{v}^j_eT(f,\hat{v}_e\cdot\hat{w}_e)\right] e^A_{ij}(\theta_e,\phi_e,\psi_e).
\end{equation}
 where all quantities are evaluated in the ecliptic coordinate system.
{\color{blue}Equation \eqref{FAecliptic} } can be used to plot the angular response functions presented in Sec. III.

\subsection{The wave frame}

\begin{figure}[htp]
  \centering
  \includegraphics[width=0.6\textwidth]{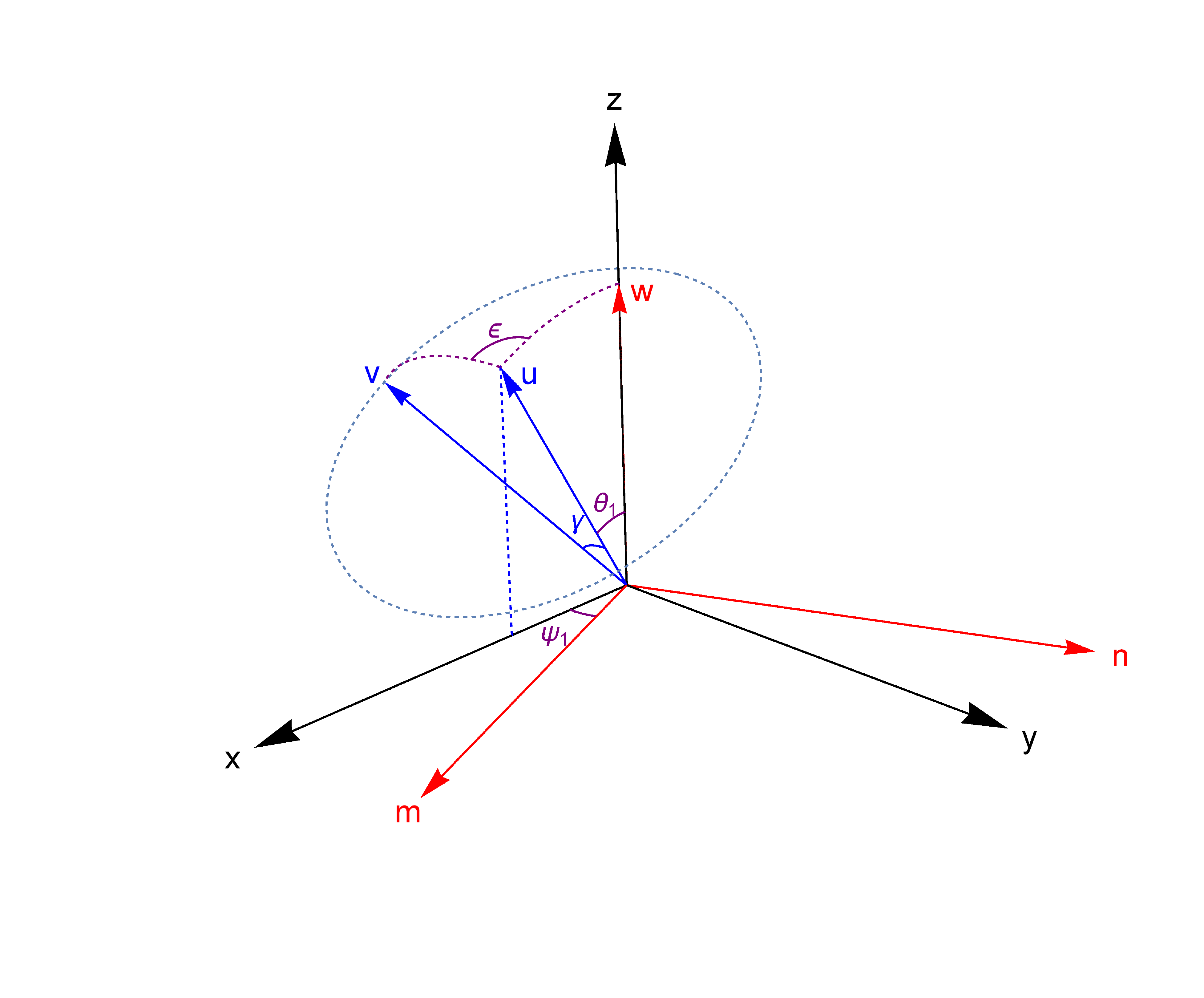}\\
  \caption{The wave coordinate system \cite{Larson:1999we}. GWs propagate along the $z$ direction.
  One arm of the detector ($\hat{u}$) locates on the $x-z$ plane with the inclination angle $\theta_1$ and the opening angle
  between the two arms $\hat{u}$ and $\hat{v}$ is $\gamma$. The angle
  between the polarization axis $\hat{m}$ and the $x$ axis is $\psi_1$, and  $\epsilon$ is the
  angle between the plane containing the vectors $\hat{u}$ and $\hat{w}$ and the plane of the interferometer.}
\label{figsource}
\end{figure}

In the wave frame \cite{Larson:1999we}, GWs propagate along the $z$ direction. The unit vector for one arm of the detector $\hat{u}$ locates on the $x-z$ plane with inclination angle $\theta_1$
and the angle
between the polarization axis $\hat{m}$ and the $x$ axis is $\psi_1$ which is measured from the point where the plane
of the first arm with the unit vector $\hat{u}$ and the propagation vector intersects the plane containing the principal polarization vector \cite{Larson:1999we}. To characterize the unit vector for the other arm of the detector, we follow \cite{Larson:1999we} to introduce $\epsilon$, which is the angle between the plane containing vector $\hat{u}$ and $\hat{w}$ and the plane of the interferometer as shown in Fig. \ref{figsource}.

Using the geometric relations in Fig. \ref{figsource}, we have
\begin{gather}
\hat{u}_w=(\sin\theta_1,0,\cos\theta_1), \\
\hat{v}_w=(\cos\gamma\sin\theta_1-\sin\gamma\cos\theta_1\cos\epsilon,-\sin\gamma\sin\epsilon,\cos\gamma\cos\theta_1+\sin\gamma\sin\theta_1\cos\epsilon), \\
\hat{w}=(0,0,1), \\
\hat{m}=(\cos\psi_1,\sin\psi_1,0), \\
\hat{n}=(-\sin\psi_1,\cos\psi_1,0).
\end{gather}
In the wave frame, the angular response function is
\begin{equation}
F^A=\frac{1}{2}\left[\hat{u}^i_w\hat{u}^j_w T(f,\cos\theta_1)- \hat{v}^i_w\hat{v}^j_w T(f,\cos\theta_2)\right] e^A_{ij}(\psi_1),
\end{equation}
where $\cos\theta_2=\cos\gamma\cos\theta_1+\sin\gamma\sin\theta_1\cos\epsilon$. So
\begin{equation}
\begin{aligned}
\label{fplus0}
F^+=\frac{1}{2}&\cos2\psi_1\sin^2\theta_1 T(f,\cos\theta_1) \\
         -\frac{1}{2}&[\cos2\psi_1((\sin\gamma\cos\theta_1\cos\epsilon-\cos\gamma\sin\theta_1)^2-\sin^2\gamma\sin^2\epsilon) \\
                      &+2\sin2\psi_1\sin\gamma\sin\epsilon(\sin\gamma\cos\theta_1\cos\epsilon-\cos\gamma\sin\theta_1) ]T(f,\cos\theta_2),
\end{aligned}
\end{equation}
\begin{equation}
\begin{aligned}
F^\times=-\frac{1}{2}&\sin2\psi_1\sin^2\theta_1 T(f,\cos\theta_1) \\
 +\frac{1}{2}&[\sin2\psi_1((\sin\gamma\cos\theta_1\cos\epsilon-\cos\gamma\sin\theta_1)^2-\sin^2\gamma\sin^2\epsilon) \\
                      &-2\cos2\psi_1\sin\gamma\sin\epsilon(\sin\gamma\cos\theta_1\cos\epsilon-\cos\gamma\sin\theta_1) ]T(f,\cos\theta_2),
\end{aligned}
\end{equation}
\begin{equation}
\begin{aligned}
F^x=\frac{1}{2}&\cos\psi_1\sin2\theta_1 T(f,\cos\theta_1) \\
   -\frac{1}{2}&(-\sin\gamma\cos\theta_1\cos\epsilon\cos\psi_1-\sin\gamma\sin\epsilon\sin\psi_1+\cos\gamma\sin\theta_1\cos\psi_1)\cos\theta_2 T(f,\cos\theta_2),
\end{aligned}
\end{equation}
\begin{equation}
\begin{aligned}
F^y=-\frac{1}{2}&\sin\psi_1\sin2\theta_1 T(f,\cos\theta_1) \\
-\frac{1}{2}&(\sin\gamma\cos\theta_1\cos\epsilon\sin\psi_1-\sin\gamma\sin\epsilon\cos\psi_1-\cos\gamma\sin\theta_1\sin\psi_1)\cos\theta_2 T(f,\cos\theta_2),
\end{aligned}
\end{equation}
\begin{equation}
\begin{aligned}
F^b=\frac{1}{2}\sin^2\theta_1 T(f,\cos\theta_1)-\frac{1}{2}(\sin^2\gamma\sin^2\epsilon+(\sin\gamma\cos\theta_1\cos\epsilon-\cos\gamma\sin\theta_1)^2) T(f,\cos\theta_2),
\end{aligned}
\end{equation}

\begin{equation}
\begin{aligned}
\label{flong0}
F^l=\frac{1}{2}\cos^2\theta_1 T(f,\cos\theta_1)-\frac{1}{2}\cos^2\theta_2 T(f,\cos\theta_2).
\end{aligned}
\end{equation}
Using Eqs. (\ref{fplus0}-\ref{flong0}) with Eq. \eqref{averagedplus}, we can obtain the results (\ref{rteq6}-\ref{rleq6}).

Since the polarization angle is $\psi_1$, the rotation matrix $R^{we}$ between wave coordinates and ecliptic coordinates is
\begin{equation}
\begin{aligned}
R^{we}=R^{wc}\cdot R^{ce}=
  \left(
    \begin{array}{cccc}
       \cos\psi_1 & \sin\psi_1 & 0 \\
       -\sin\psi_1 & \cos\psi_1 & 0 \\
       0 & 0 & 1  \\
    \end{array}
  \right)\cdot R^{ce},
\end{aligned}
\end{equation}

\subsection{The detector frame}

\begin{figure}[htp]
  \centering
  \includegraphics[width=0.6\textwidth]{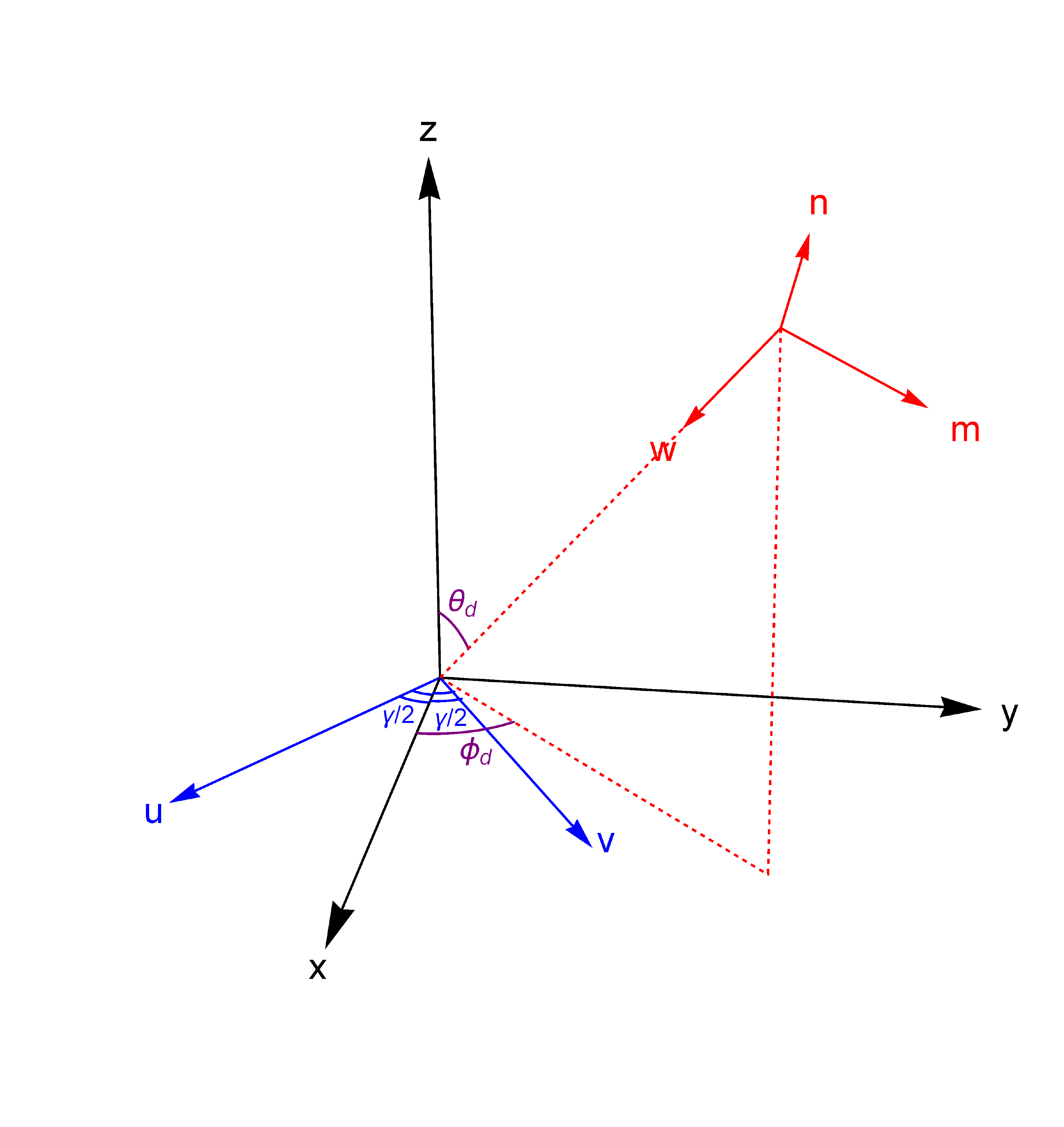}
  \caption{The detector coordinate system \cite{Guersel:1989th,Giampieri:1997kv}. The two arms of the detector lie along the direction $(\cos(\gamma/2),-\sin(\gamma/2),0)$
  and $(\cos(\gamma/2),\sin(\gamma/2),0)$, and the source is located at the direction $-\hat{w}=(\sin\theta_d\cos\phi_d,\sin\theta_d\sin\phi_d,\cos\theta_d)$.}
\label{figdetector}
\end{figure}

The detector coordinate system is chosen such that the two arms of the detector keep fixed
on the $x-y$ plane with one lying along the direction $(\cos(\gamma/2),-\sin(\gamma/2),0)$
and the other lying along the direction $(\cos(\gamma/2),\sin(\gamma/2),0)$ \cite{Guersel:1989th,Giampieri:1997kv},
here $\gamma$ is the opening angle between the two arms of the detector.
The source is located in the direction $(\theta_d,\phi_d)$.
The detector coordinate system is shown in Fig. \ref{figdetector}.
So
\begin{gather}
\hat{u}_d=(\cos(\gamma/2),-\sin(\gamma/2),0), \\
\hat{v}_d=(\cos(\gamma/2),\sin(\gamma/2),0).
\end{gather}
In the detector frame, the angular response function is
\begin{equation}
F^A=\frac{1}{2}\left[\hat{u}^i_d\hat{u}^j_d T(f,\hat{u}_d\cdot\hat{w}_d)- \hat{v}^i_d\hat{v}^j_dT(f,\hat{v}_d\cdot\hat{w}_d)\right] e^A_{ij}(\theta_d,\phi_d,\psi).
\end{equation}
So
\begin{equation}
\label{fplus1}
\begin{split}
F^+&=-\frac{\sin\gamma}{4}(T_u+T_v)[(1+\cos^2\theta_d)\sin2\phi_d\cos2\psi+2\cos\theta_d\cos2\phi_d\sin2\psi] \\
   &+\frac{\cos^2(\gamma/2)}{2}(T_u-T_v)[(\cos^2\theta_d\cos^2\phi_d-\sin^2\phi_d)\cos2\psi-\cos\theta_d\sin2\phi_d\sin2\psi] \\
   &+\frac{\sin^2(\gamma/2)}{2}(T_u-T_v)[(\cos^2\theta_d\sin^2\phi_d-\cos^2\phi_d)\cos2\psi+\cos\theta_d\sin2\phi_d\sin2\psi],
\end{split}
\end{equation}
\begin{equation}
\label{fcross1}
\begin{split}
F^\times&=-\frac{\sin\gamma}{4}(T_u+T_v)[-(1+\cos^2\theta_d)\sin2\phi_d\sin2\psi+2\cos\theta_d\cos2\phi_d\cos2\psi] \\
     &+\frac{\cos^2(\gamma/2)}{2}(T_u-T_v)[-(\cos^2\theta_d\cos^2\phi_d-\sin^2\phi_d)\sin2\psi-\cos\theta_d\sin2\phi_d\cos2\psi] \\
   &+\frac{\sin^2(\gamma/2)}{2}(T_u-T_v)[-(\cos^2\theta_d\sin^2\phi_d-\cos^2\phi_d)\sin2\psi+\cos\theta_d\sin2\phi_d\cos2\psi],
\end{split}
\end{equation}
\begin{equation}
\label{fvecx1}
\begin{split}
F^x&=-\frac{\sin\gamma}{4}(T_u+T_v)[\sin2\theta_d\sin2\phi_d\cos\psi+2\sin\theta_d\cos2\phi_d\sin\psi] \\
     &+\frac{\cos^2(\gamma/2)}{2}(T_u-T_v)[\sin2\theta_d\cos^2\phi_d\cos\psi-\sin\theta_d\sin2\phi_d\sin\psi] \\
   &+\frac{\sin^2(\gamma/2)}{2}(T_u-T_v)[\sin2\theta_d\sin^2\phi_d\cos\psi+\sin\theta_d\sin2\phi_d\sin\psi],
\end{split}
\end{equation}
\begin{equation}
\label{fvecy1}
\begin{split}
F^y&=-\frac{\sin\gamma}{4}(T_u+T_v)[-\sin2\theta_d\sin2\phi_d\sin\psi+2\sin\theta_d\cos2\phi_d\cos\psi] \\
     &+\frac{\cos^2(\gamma/2)}{2}(T_u-T_v)[-\sin2\theta_d\cos^2\phi_d\sin\psi-\sin\theta_d\sin2\phi_d\cos\psi] \\
   &+\frac{\sin^2(\gamma/2)}{2}(T_u-T_v)[-\sin2\theta_d\sin^2\phi_d\sin\psi+\sin\theta_d\sin2\phi_d\cos\psi],
\end{split}
\end{equation}
\begin{equation}
\label{fbreath1}
\begin{split}
F^b&=\frac{\sin\gamma}{4}(T_u+T_v)\sin^2\theta_d\sin2\phi_d
     +\frac{\cos^2(\gamma/2)}{2}(T_u-T_v)(\cos^2\theta_d\cos^2\phi_d+\sin^2\phi_d) \\
   &+\frac{\sin^2(\gamma/2)}{2}(T_u-T_v)(\cos^2\theta_d\sin^2\phi_d+\cos^2\phi_d),
\end{split}
\end{equation}
\begin{equation}
\label{flong1}
\begin{split}
F^l&=-\frac{\sin\gamma}{4}(T_u+T_v)\sin^2\theta_d\sin2\phi_d
     +\frac{\cos^2(\gamma/2)}{2}(T_u-T_v)\sin^2\theta_d\cos^2\phi_d \\
   &+\frac{\sin^2(\gamma/2)}{2}(T_u-T_v)\sin^2\theta_d\sin^2\phi_d,
\end{split}
\end{equation}
where $T_u=T(f,-\sin\theta_d\cos(\phi_d+\gamma/2))$
and $T_v=T(f,-\sin\theta_d\cos(\phi_d-\gamma/2))$.

The unit vectors $\{\hat{m},\hat{n},\hat{w}\}$ is related to the detector coordinates $\{x_d,y_d,z_d\}$
by the Euler rotation matrix
\begin{equation}
\begin{aligned}
R_{ij}^{wd}=
  \left(
    \begin{array}{cccc}
   \cos\theta_d\cos\phi_d\cos\psi+\sin\phi_d\sin\psi & -\cos\theta_d\cos\phi_d\sin\psi+\sin\phi_d\cos\psi &  -\sin\theta_d\cos\phi_d    \\
   \cos\theta_d\sin\phi_d\cos\psi-\cos\phi_d\sin\psi & -\cos\theta_d\sin\phi_d\sin\psi-\cos\phi_d\cos\psi & -\sin\theta_d\sin\phi_d \\
       -\sin\theta_d\cos\psi & \sin\theta_d\sin\psi & -\cos\theta_d     \\
    \end{array}
  \right).
\end{aligned}
\end{equation}
The rotation matrix $R^{wd}$ between the detector coordinates and the wave coordinates is
\begin{equation}
R^{wd}=R^{wc}\cdot R^{cd}.
\end{equation}
By using this coordinate transformation, we can derive Eqs. (\ref{fplus0}-\ref{flong0}) from  Eqs. (\ref{fplus1}-\ref{flong1}).

Now, let us consider the relationship between the detector coordinates and the ecliptic coordinates for TianQin and LISA.
For TianQin, its detector coordinates $\{x_t,y_t,z_t\}$ is related to the ecliptic coordinates $\{x_e,y_e,z_e\}$ by the rotation matrix
\begin{equation}
\begin{aligned}
&R_{ij}^{te}= \\
 & \left(
    \begin{array}{cccc}
   \cos\theta_{tq}\cos\phi_{tq}\sin\alpha_{ot}+\sin\phi_{tq}\cos\alpha_{ot} & \cos\theta_{tq}\cos\phi_{tq}\cos\alpha_{ot}-\sin\phi_{tq}\sin\alpha_{ot} &  \sin\theta_{tq}\cos\phi_{tq}    \\
   \cos\theta_{tq}\sin\phi_{tq}\sin\alpha_{ot}-\cos\phi_{tq}\cos\alpha_{ot} & \cos\theta_{tq}\sin\phi_{tq}\cos\alpha_{ot}+\cos\phi_{tq}\sin\alpha_{ot} & \sin\theta_{tq}\sin\phi_{tq} \\
       -\sin\theta_{tq}\sin\alpha_{ot} & -\sin\theta_{tq}\cos\alpha_{ot} & \cos\theta_{tq}     \\
    \end{array}
  \right),
\end{aligned}
\end{equation}
where $\alpha_{ot}=2\pi f_{sc} t+\pi-\alpha_{T0}'$ and $\alpha_{T0}'$ is the initial phase.
We can also obtain the relationship between the source location $(\theta_t,\phi_t)$ in detector coordinates
and $(\theta_e,\phi_e)$  in ecliptic coordinates for TianQin by using
$$(\sin\theta_t\cos\phi_t,\sin\theta_t\sin\phi_t,\cos\theta_t)=(R^{te})^{-1} \cdot (\sin\theta_e\cos\phi_e,\sin\theta_e\sin\phi_e,\cos\theta_e),$$
which tells us how an arbitrary source moves with time in detector coordinates. We can also verify that
$R^{te}\cdot \hat{u}_d=\hat{u}_e$ and $R^{te}\cdot \hat{v}_d=\hat{v}_e$.

For LISA, the detector coordinates $\{x_l,y_l,z_l\}$ is related to the ecliptic coordinates $\{x_e,y_e,z_e\}$ by the rotation matrix
\begin{equation}
\begin{aligned}
&R_{ij}^{le}= \\
 & \left(
    \begin{array}{cccc}
   \cos\beta\cos^2\alpha_{ol}+\sin^2\alpha_{ol} & (\cos\beta-1)\sin\alpha_{ol}\cos\alpha_{ol} &  -\sin\beta\cos\alpha_{ol}    \\
  (\cos\beta-1)\sin\alpha_{ol}\cos\alpha_{ol} & \cos\beta\sin^2\alpha_{ol}+\cos^2\alpha_{ol} & -\sin\beta\sin\alpha_{ol} \\
       \sin\beta\cos\alpha_{ol} & \sin\beta\sin\alpha_{ol} & \cos\beta     \\
    \end{array}
  \right),
\end{aligned}
\end{equation}
where $\alpha_{ol}=2\pi f_m t-\alpha_{L0}$, and $\beta=\pi/3$ is the inclination of the detector plane to the ecliptic plane,
$\alpha_{L0}$ is the initial phase.
Thus, the rotation matrices $R^{tw}$ and $R^{lw}$ between the detector frame and the wave frame for TianQin and LISA are
\begin{equation}
R^{tw}=R^{te}\cdot(R^{we})^{-1},
\end{equation}
\begin{equation}
R^{lw}=R^{le}\cdot(R^{we})^{-1}.
\end{equation}


%

\end{document}